\documentclass[conference]{IEEEtran}
\IEEEoverridecommandlockouts

\usepackage{booktabs}
\usepackage{graphicx}
\usepackage{cite}
\usepackage{amsmath,amssymb,amsfonts}
\usepackage{algorithm}
\usepackage{algpseudocode}
\usepackage{textcomp}
\usepackage[dvipsnames]{xcolor}
\usepackage{csquotes}
\usepackage{multirow}
\usepackage{hyperref}
\usepackage{balance}
\usepackage{caption}
\usepackage{enumitem}
\usepackage{microtype}
\usepackage{tikz}
\usepackage{tabularx}
\usetikzlibrary{arrows.meta,positioning,fit,backgrounds,calc}
\usepackage{mdframed}

\newmdenv[
  backgroundcolor=black!4,
  linecolor=black!50,
  linewidth=0.5pt,
  innertopmargin=4pt,
  innerbottommargin=4pt,
  innerleftmargin=6pt,
  innerrightmargin=6pt,
  skipabove=4pt,
  skipbelow=4pt,
]{ResultBox}

\newif\ifcomments
\commentsfalse

\ifcomments
  \newcommand{\song}[1]{{\textcolor{orange}{\textbf{Song:}~\enquote{#1}}}}
  \newcommand{\amal}[1]{{\textcolor{Mahogany}{\textbf{Amal:}~\enquote{#1}}}}
  \newcommand{\kyle}[1]{{\textcolor{Magenta}{\textbf{Kyle:}~\enquote{#1}}}}
  \newcommand{\ian}[1]{{\textcolor{olive}{\textbf{Ian:}~\enquote{#1}}}}
  \newcommand{\fm}[1]{{\textcolor{red}{\textbf{FIXME:}~\enquote{#1}}}}
\else
  \newcommand{\song}[1]{}
  \newcommand{\amal}[1]{}
  \newcommand{\kyle}[1]{}
  \newcommand{\ian}[1]{}
  \newcommand{\fm}[1]{}
\fi

\def\BibTeX{{\rm B\kern-.05em{\sc i\kern-.025em b}\kern-.08em
    T\kern-.1667em\lower.7ex\hbox{E}\kern-.125emX}}

\newcommand{\compass}{\textsc{Compass}}
\newcommand{\prikg}{\textsf{PriKG}}
\newcommand{\hipkg}{\textsf{HipKG}}
\newcommand{\phakg}{\textsf{PhaKG}}
\newcommand{\gokg}{\textsf{GoKG}}

\begin{document}

\title{\compass{}: Steering Distributed Vector Search\\with Scientific Knowledge Graphs}

\author{
    \IEEEauthorblockN{
        Song Young Oh\IEEEauthorrefmark{1},
        Amal Gueroudji\IEEEauthorrefmark{2},
        Seth Ockerman\IEEEauthorrefmark{3},
        Rob Latham\IEEEauthorrefmark{2},\\
        Orcun Yildiz\IEEEauthorrefmark{2},
        Ian Foster\IEEEauthorrefmark{4}\IEEEauthorrefmark{1},
        Kyle Chard\IEEEauthorrefmark{1}\IEEEauthorrefmark{4},
        Robert Ross\IEEEauthorrefmark{2}
    }
    \IEEEauthorblockA{
        \IEEEauthorrefmark{1}Department of Computer Science, University of Chicago, Chicago, IL, USA\\
    }
    \IEEEauthorblockA{
        \IEEEauthorrefmark{2}Mathematics and Computer Science Division, Argonne National Laboratory, Lemont, IL, USA\\
    }
    \IEEEauthorblockA{
        \IEEEauthorrefmark{3}Department of Computer Science, University of Wisconsin--Madison, Madison, WI, USA\\
    }
    \IEEEauthorblockA{
        \IEEEauthorrefmark{4}Data Science and Learning Division, Argonne National Laboratory, Lemont, IL, USA\\
    }
}

\maketitle

\begin{abstract}
Vector databases use hashing to partition data across ``shards,'' logical units for distributed execution. This placement, however, destroys semantic locality, forcing each query into scatter--gather limited by the slowest shard. Vector-space clustering can help, but scientific evidence is often connected by factual relations that do not align with embedding distance. We present \compass{}, a framework that uses a knowledge graph (KG) to determine data placement and query-time shard selection. \compass{} detects communities, splits oversized communities, inserts embeddings by subject entity, and routes queries to a small set of shards. Across four biomedical KGs, our method searches only 13--18\% of the corpus while preserving broadcast recall and recovering up to 2.6$\times$ more multi-hop evidence than an embedding-based baseline. On 15 HPC nodes, \compass{} sustains 7.9$\times$ higher throughput with lower tail latency than hash-based broadcast. These results show that KG structure provides a compact complement to embedding geometry for scalable vector search.
\end{abstract}


\section{Introduction}
Vector search has become an essential data service for artificial intelligence (AI) agents in scientific research~\cite{echihabi2021new, gueroudji2026beyond, oh2025smurf}. As collections grow, vector databases divide their data across shards to increase capacity~\cite{ockerman2025exploring, abdelhafiz2020distributed, shethiya2025load}. Most systems use hash-based methods to split data into shards because hashing is computationally inexpensive and provides balanced data placement~\cite{wang2021milvus, solat2024sharding}. The resulting layout, however, provides no semantic locality: evidence relevant to one question may be scattered across different shards. Each query must therefore search every shard and wait for a full scatter--gather to complete. Adding shards thus increases coordination overhead and total per-query cost \cite{ockerman2026more}.

An alternative to scatter--gather is selective search, which routes each query to only a subset of shards. Inverted File (IVF) indexes utilize selective search by partitioning vectors into $k$-means clusters and limiting search to only the $nProbe$ ``closest" clusters, defined by distance between each cluster's centroid and the query~\cite{zobel2006inverted, jegou2011product}. A distributed vector database can similarly co-locate related vectors and search only the most relevant shards. In scientific workloads, however, relevance often depends on structured relationships that embedding similarity does not reliably preserve~\cite{auer2018towards, peng2023knowledge, mohamed2021biological}. For example, a drug, its target protein, and a disease mechanism can be linked by typed relations despite distant textual embeddings~\cite{vittor2026optimuskg}. As a result, embedding-based placement may scatter evidence needed for the same queries across too many shards.

\begin{figure}[t]
\centering
\includegraphics[width=0.95\columnwidth]{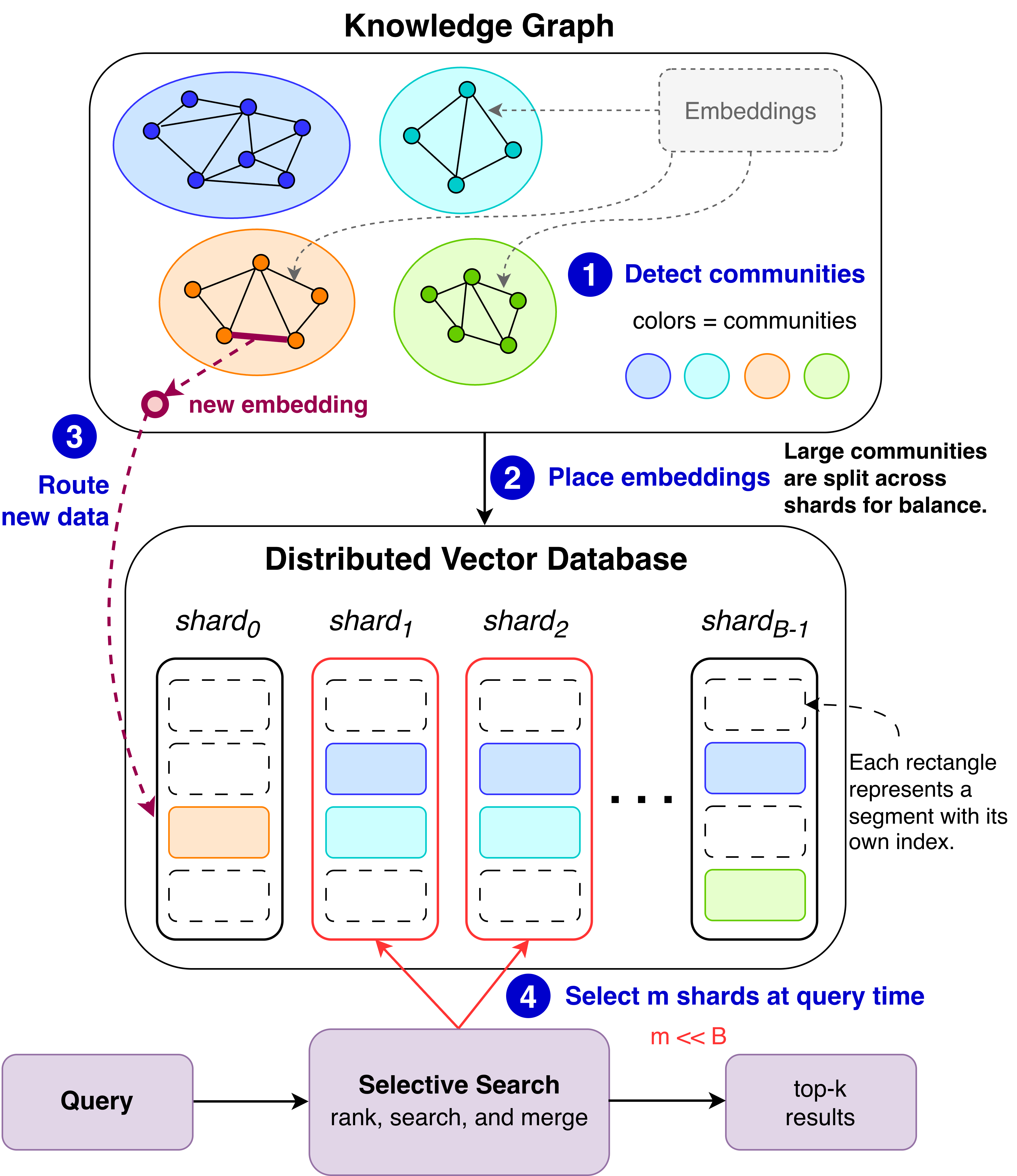}
\caption{\compass{} overview. \compass{} detects graph communities and splits oversized communities to balance shard sizes. It assigns new vectors to shards based on their graph neighborhoods and searches only $m\ll B$ shards per query.}
\label{fig:overview}
\end{figure}

Knowledge graphs (KGs) provide a complementary placement signal by explicitly encoding domain-specific relationships that embeddings may miss~\cite{mohamed2025knowledge}. Using this signal for sharding requires deciding how to partition the graph while preserving useful relationships. Graph systems typically address this with edge-cut methods that balance vertices, while vertex-cut and replication schemes further handle high-degree vertices and load skew~\cite{karypis1998fast, stanton2012streaming, tsourakakis2014fennel, gonzalez2012powergraph}. For semantic search, KG communities---densely connected groups of entities---offer a natural basis for co-locating associated vectors likely to support related or multi-hop queries~\cite{traag2019louvain}. However, community sizes can vary substantially~\cite{de2011generalized}, so placement must balance graph locality against shard load.

We present \compass{} (\textbf{COM}munity-guided \textbf{P}lacement \textbf{A}nd \textbf{S}hard \textbf{S}election), a data placement and routing layer for distributed vector databases (\autoref{fig:overview}). During the initial upload phase, \compass{} detects KG communities, splits only those too large for balanced placement, and assigns the resulting pieces to shards. Embeddings are stored with their subject entities: those for known entities follow the existing entity-to-shard map, while those for unseen entities are placed based on their graph neighborhoods. At query time, personalized PageRank (PPR)~\cite{haveliwala2003analytical} scores shards from entities named in the query, and shard-local approximate nearest-neighbor (ANN) indexes search only the highest-ranked shards~\cite{arya1998optimal}.

Our evaluation separates the system's benefit of selective search from the quality benefit of graph-guided insertion. With the same two-shard search, \compass{} and $k$-means achieve similar query throughput and latency, but \compass{} preserves broadcast recall and retrieves up to $2.6\times$ as much multi-hop evidence. It raises throughput from 9.0k to 71.8k queries/s and reduces p95 latency from 8.1 to 1.0 ms over broadcast on 15 HPC nodes. Further analysis shows that these gains come from co-locating related data rather than exploiting answer-specific graph links during routing.

This work makes the following three contributions:
\begin{enumerate}[leftmargin=1.45em,itemsep=1pt,topsep=2pt]
\item We formulate distributed vector search as a joint locality and load-balance problem, demonstrating that embedding proximity alone may not preserve relational evidence.

\item We design and implement \compass{}, which combines bounded KG communities, graph-based online placement, and selective shard search.

\item We evaluate \compass{} on four biomedical KGs, measuring retrieval quality, system performance, and the conditions under which graph-aware placement is most effective.
\end{enumerate}

\section{Related Work}
\label{sec:related}

\textbf{Distributed vector and graph systems.}
Sharding, a standard technique in distributed databases, partitions storage and computation across machines to improve scalability and availability~\cite{abdelhafiz2020distributed, solat2024sharding, shethiya2025load}. Vector databases adopt this design: each shard stores a subset of the vectors in one or more segments, with each segment maintaining its own index~\cite{ma2023comprehensive}. Widely used systems, such as Qdrant~\cite{qdrant2025} and Milvus~\cite{wang2021milvus}, assign vectors to shards by hashing their identifiers, often through consistent hashing~\cite{karger1997consistent}. Because this approach does not preserve semantic relationships, queries generally follow a scatter--gather model: they search every shard and merge the local results~\cite{pan2024survey, ockerman2026more}.
% Shard-level placement is distinct from organization within each local index. Indexing methods such as IVF~\cite{zobel2006inverted} and IVF-PQ~\cite{jegou2011product} use $k$-means internally but do not coordinate data placement across shards. \compass{} instead uses knowledge graph structure to guide shard-level placement as vectors are inserted.

Distributed graph processing systems also divide vertices and edges while preserving locality. Classical edge-cut methods map vertices to balanced partitions and minimize cross-partition edges~\cite{karypis1998fast, stanton2012streaming, tsourakakis2014fennel}. For graphs with skewed degree distributions, vertex-cut systems such as PowerGraph~\cite{gonzalez2012powergraph} instead partition edges and replicate vertices whose incident edges span multiple partitions, particularly high-degree vertices~\cite{ozsu2016survey, huang2011scalable}. \compass{} adopts this locality--balance objective for vector placement.

\textbf{Graph-guided retrieval.}
Graph RAG methods use graph links to retrieve connected evidence~\cite{gutierrez2024hipporag,wu2024medical,han2024retrieval, li2025simple}, traverse reasoning paths~\cite{sun2023think,luo2024reasoning}, or build lightweight entity--relation indexes~\cite{guo2024lightrag}. This literature mostly focuses on which evidence to return, while treating the vector store layout as fixed~\cite{peng2025graph}. Closer to our distributed setting, DGRAG~\cite{zhou2025dgrag} builds local KGs at edge devices, summarizes their subgraphs, and uses a cloud coordinator to route queries to relevant devices. However, its partitions are determined by data ownership, whereas \compass{} derives shard boundaries from knowledge graph structure.

To form these boundaries, \compass{} builds on community detection, which identifies densely connected regions of a graph~\cite{newman2004finding}. Louvain~\cite{blondel2008fast} and Leiden~\cite{traag2019louvain} optimize modularity, which favors partitions containing more within-group edges than expected under a null model~\cite{newman2006modularity}, while alternatives include label propagation, spectral clustering, and stochastic block models~\cite{raghavan2007near,rosvall2008maps,von2007tutorial,karrer2011stochastic}. Because these methods do not enforce size constraints and can yield highly uneven communities~\cite{fortunato2016community}, \compass{} treats their outputs as candidate placement units and recursively splits only those that exceed a bound.

\section{Methodology}
\label{sec:method}

\subsection{Overview}

Let $G=(V,E)$ be a knowledge graph, $F$ a corpus of embedded facts, and $B$ the number of vector database shards. Each fact $f\in F$ has a subject entity $s(f)\in V$. Given a query $q$, the router selects $m$ shards, searches their local ANN indexes, and merges the resulting top-$k$ candidates. Facts and entities share an embedding space; $e_x$ denotes the embedding associated with entity $x$, $e_q$ the query embedding, and $\mu_b$ the centroid of shard $b$'s fact embeddings.

Selective search is effective only when relevant facts concentrate on a small number of shards and shard sizes stay balanced. \compass{} addresses both through a shared entity-to-shard map $\beta:V\rightarrow\{1,\ldots,B\}$, which records where facts are stored and how graph evidence becomes shard-level routing decisions. Query work is therefore proportional to the facts on the selected shards,
\begin{equation}
W(q)=\sum_{b\in S(q)}|F_b|,
\label{eq:work}
\end{equation}
and latency is determined by routing, the slowest selected shard, and result merging.

\compass{} maintains this mapping across three phases: offline data placement, online insertion of new facts, and query-time routing to $m$ shards. Algorithm~\ref{alg:compass} summarizes these phases and the rest of this section presents the details.

\subsection{Data placement}
\label{sec:placement}

We construct the initial layout from a bootstrap graph
$G_{\mathrm{bulk}}=(V_{\mathrm{bulk}},E_{\mathrm{bulk}})$. \compass{} applies Louvain community detection~\cite{de2011generalized}, which greedily groups neighboring vertices to increase modularity, a measure of how strongly vertices connect within communities relative to across them. It then collapses each community into a supernode and repeats until modularity no longer improves. A large community can overload one shard, so we bound each piece by $C_{\max}=\lceil\rho\,|V_{\mathrm{bulk}}|/B\rceil$, where $\rho$ controls the maximum piece size relative to the average shard load. Communities above the bound are recursively re-partitioned by Louvain on their induced subgraphs, and smaller ones remain intact.

Let $\mathcal{C}$ denote the resulting pieces. For piece $C$, $h_C$ is its normalized entity-type histogram and $|C|$ its size; for shard $b$, $h_b$ and $L_b$ denote its type histogram and entity load. Pieces are processed largest first: the first $B$ seed empty shards, and each remaining piece is assigned to the shard maximizing
\begin{equation}
\mathrm{place}(C,b)=\cos(h_C,h_b)-\lambda,\frac{L_b}{|V_{\mathrm{bulk}}|/B},
\label{eq:place}
\end{equation}
where $\lambda$ balances type compatibility against shard load. All entities in a piece share its shard assignment, and each fact is stored with its subject entity. Thus, the KG determines placement across shards, while each local ANN index ranks facts by embedding similarity.

\begin{algorithm}[t]
\caption{\compass{} data placement and query routing}
\label{alg:compass}
\small
\begin{algorithmic}[1]
\setlength{\itemsep}{1.5pt}
\Require bootstrap graph $G_{\mathrm{bulk}}$, facts $F$, shards $B$

\Statex \emph{Offline placement}\quad
($\mathrm{place}$: Eq.~\ref{eq:place})
\State $G_u \gets \Call{Undirected}{G_{\mathrm{bulk}}}$
\State $\mathcal{C} \gets \Call{Louvain}{G_u}$
\State split each $C\in\mathcal{C}$ with
$|C|>C_{\max}=\lceil\rho|V_{\mathrm{bulk}}|/B\rceil$

\For{$C\in\mathcal{C}$ in decreasing order of $|C|$}
  \State $b^\star\gets$ empty shard if any; otherwise,
  $\operatorname*{arg\,max}_{b}\mathrm{place}(C,b)$
  \State $\beta(v)\gets b^\star\ \forall v\in C$;
  update $L_{b^\star},h_{b^\star}$
\EndFor

\State store each fact $f\in F$ on shard $\beta(s(f))$

\Statex \emph{Online routing}\quad
($\mathrm{score}$: PPR mass per shard, Eq.~\ref{eq:score})
\State \textbf{insert} unseen $x$: shard
$\operatorname*{arg\,max}_{b}\mathrm{score}(N(x),b)$;
otherwise, nearest $\mu_b$
\State \textbf{query} $q$: top-$m$ shards by
$\mathrm{score}(A_q,b)$; otherwise, by $\cos(e_q,\mu_b)$
\end{algorithmic}
\end{algorithm}

\subsection{Online insertion}
\label{sec:insertion}

New entities subsequently arrive as an insertion stream. A fact whose subject is already known is sent directly to $\beta(s(f))$. Both insertion and query routing use personalized PageRank (PPR)~\cite{haveliwala2003analytical}: from an anchor set $A$, PPR yields a distribution $\pi_A$ over entities, and each shard is scored by the mass on its assigned entities,
\begin{equation}
\mathrm{score}(A,b)=\sum_{v:\beta(v)=b}\pi_A(v).
\label{eq:score}
\end{equation}
PPR uses restart probability $\alpha$ and sparse power iteration, keeping mass near the anchors while allowing propagation through indirect relationships.

For an unseen entity $x$, \compass{} uses its known neighbors $N(x)$ as anchors and assigns $x$ to the shard maximizing $\mathrm{score}(N(x),b)$; later facts about $x$ inherit this assignment. If $x$ has no known neighbor, it is placed at the nearest shard centroid $\mu_b$.

\subsection{Query routing}
\label{sec:routing}

At query time, \compass{} uses the same graph-based shard scoring. A conservative entity linker extracts only entities named in $q$ as anchors $A_q$. \compass{} runs PPR, searches the $m$ highest-scoring shards, and merges their top-$k$ results. If no entity is linked, shards are instead ranked by similarity between the query embedding $e_q$ and shard centroids $\mu_b$.

The router stores only the entity-to-shard map, truncated graph adjacency, and one centroid per shard. It operates outside the vector database and requires no changes to local ANN indexing or the query interface.

\section{Experimental Setup}
\label{sec:experiments}

\subsection{Datasets and metrics}

We evaluate \compass{} on four biomedical KGs (\autoref{tab:datasets}) used in SciCUEval~\cite{yu2025scicueval}: PrimeKG~\cite{chandak2023building} (\prikg{}), HIPPIE~\cite{alanis2016hippie} (\hipkg{}), PharmKG~\cite{zheng2021pharmkg} (\phakg{}), and Gene Ontology~\cite{ashburner2000gene} (\gokg{}). For each KG, we deduplicate and combine ground-truth facts with distractors so all methods search the same corpus. SciCUEval classifies questions as (i) relevant information identification, (ii) information integration, or (iii) context-aware inference. We group the latter two as \emph{multi-hop} because both require combining evidence across entities.

Facts and questions are embedded with the 768-dimensional
\textsf{S-PubMedBert-MS-MARCO} encoder. We primarily report \emph{fact recall}@10, the fraction of a question's ground-truth facts returned in the top 10; ``recall'' denotes this throughout. It scores which facts \emph{answer} a question (semantic ground-truth), not which vectors are nearest to it (ANN ground-truth). We also report gold-shard coverage, whether the selected shards hold the evidence before local ranking; corpus fraction searched, which accounts for unequal shard sizes; shard-load Gini, the imbalance of facts across shards ($0$ uniform, $1$ maximally skewed); query throughput and p50/p95/p99 latency; and router memory. Held-out edge recovery and RAG answer accuracy provide additional measures of embedding quality and end-to-end utility.

\begin{table}[t]
\centering
\caption{Biomedical knowledge graph datasets.}
\label{tab:datasets}
\small
\setlength{\tabcolsep}{6.0pt}
\begin{tabular}{@{}lrrr@{}}
\toprule
KG & Edges (facts) & Nodes (entities) & Questions \\
\midrule
\prikg{} & 298k & 54k & 1044 \\
\hipkg{} & 190k & 17k & 909 \\
\phakg{} & 91k  & 43k & 961 \\
\gokg{}  & 44k  & 44k & 926 \\
\bottomrule
\end{tabular}
\end{table}

\subsection{Baselines and system}

We compare \compass{} against two baselines. (1)~\emph{ID-hash (broadcast)} assigns facts to shards by identifier and searches all shards for every query. (2)~\emph{$k$-means} clusters fact embeddings, assigns each cluster to a shard, and routes queries to the nearest cluster centroids. It uses the same shard budget as \compass{} but relies only on embedding similarity rather than graph structure.

We further evaluate two alternative placement strategies to isolate the effect of bounded community splitting. (1)~\emph{No community split} keeps each Louvain community intact and assigns it to a single shard, preserving graph locality but allowing oversized communities to create load imbalance. (2)~\emph{Fully scattered} distributes the members of each community across shards, improving load balance while eliminating community locality.

Experiments use Qdrant v1.16~\cite{qdrant2025} with a Rust gRPC client on Aurora supercomputer at Argonne Leadership Computing Facility. Each shard uses a local HNSW index with \texttt{m}=16 and \texttt{ef}=128, stored in node-local \texttt{tmpfs}.

\begin{figure*}[t]
\centering
\includegraphics[width=\textwidth]{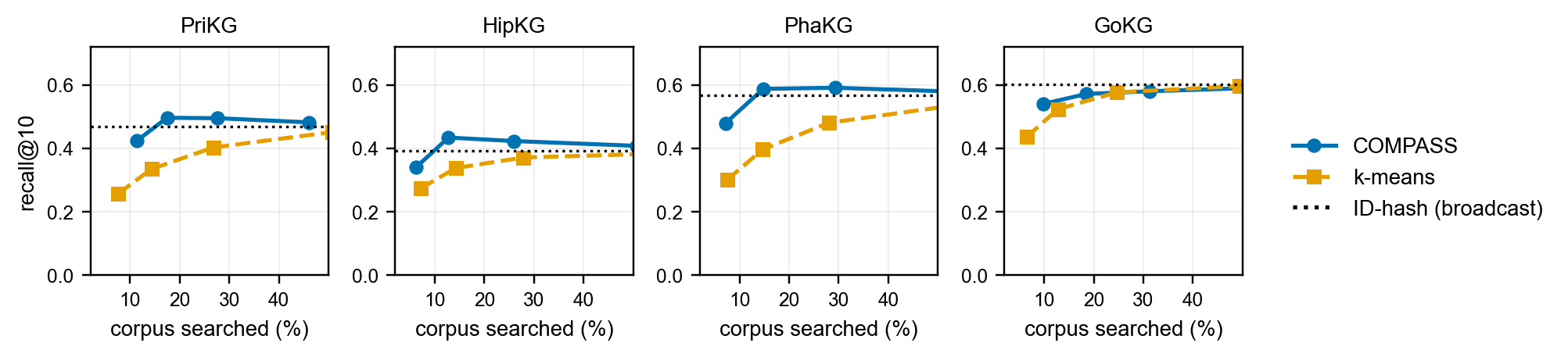}
\caption{Recall@10 versus corpus fraction searched at 16 shards. Each curve varies the number of probed shards; the dotted line marks broadcast recall. \compass{} consistently reaches higher recall than $k$-means at the same search cost and matches broadcast's performance while searching less than 20\% of the corpus.}
\label{fig:recall}
\end{figure*}

\subsection{Protocol}

We evaluate retrieval quality on all four KGs using 16 shards, with selective methods probing two shards by default. \compass{} uses $\rho=1.5$, $\lambda=0.5$, $\alpha=0.5$, 7 PPR iterations, and at most 128 outgoing neighbors per entity unless specified otherwise. For system-level experiments, we use \prikg{}, the largest corpus, varying the shard count from 32 to 128, scaling a 60-shard deployment from 1 to 15 nodes, and varying HNSW segments and Qdrant workers.

\subsection{Answer-leakage controls}

Because the indexed facts and routing graph come from the same KGs, we use three mechanisms to prevent answer information from influencing routing. (1)~Initial placement uses a 70\% bootstrap graph; all remaining entities are inserted online. (2)~At query time, the router uses only entities explicitly mentioned in the question, excluding answer entities and intermediate entities. (3)~After placement is fixed, we remove the ground-truth evidence edges for each query before running PPR. These controls test whether improvements come from the persistent data layout rather than the direct traversal of answer edges (\autoref{sec:quality}).

\section{Results}
\label{sec:results}

\subsection{Selective search preserves recall}
\label{sec:quality}

\autoref{fig:recall} compares recall@10 with the corpus fraction searched at 16 shards. Across all four KGs, \compass{} reaches a given recall while searching no more data than $k$-means. With two probed shards (\autoref{tab:main}), it searches 12.8--18.5\% of the corpus, matches or exceeds broadcast recall on \prikg{}, \hipkg{}, and \phakg{}, and on \gokg{} reaches 0.571, within 0.03 of broadcast's 0.601. The smaller gain on \gokg{} is consistent with its embeddings already placing many graph-connected entities close together (\autoref{sec:value}).

The largest gains occur on multi-hop questions, for which \compass{} retrieves 2.6$\times$, 1.8$\times$, and 2.0$\times$ more relevant evidence than $k$-means on \prikg{}, \hipkg{}, and \phakg{}, respectively.  For these questions, relevant facts may be linked in the graph yet distant in embedding space. As a result, $k$-means may place them on different shards, whereas \compass{} tends to co-locate them. Our selective search occasionally exceeds broadcast because it filters out high-scoring but irrelevant results; we treat these cases as preserving, rather than improving upon, broadcast quality.

To rule out answer-edge leakage, we fix the placement and remove each query's ground-truth edges before running PPR. We observe that gold-shard coverage changes by at most 0.028, and routing from question entities alone selects nearly the same shards. The gains therefore come from persistent co-location, not direct traversal of answer edges.

\begin{table}[t]
\centering
\caption{Recall@10 with 2 of 16 shards selected for search.}
\label{tab:main}

{\normalsize
\setlength{\tabcolsep}{4.1pt}
\renewcommand{\arraystretch}{1.22}

\resizebox{\columnwidth}{!}{%
\begin{tabular}{@{}lccccccc@{}}
\toprule
&
\multicolumn{3}{c}{Overall recall@10}
&
\multicolumn{1}{c}{Corpus searched}
&
\multicolumn{3}{c}{Multi-hop recall@10}
\\
\cmidrule(lr){2-4}
\cmidrule(lr){6-8}
KG
& \textbf{\compass{}}
& $k$-means
& Broadcast
& Fraction
& \textbf{\compass{}}
& $k$-means
& Broadcast
\\
\midrule
\prikg{}
& \textbf{0.496}
& 0.335
& 0.469
& 17.4\%
& \textbf{0.324}
& 0.124
& 0.246
\\
\hipkg{}
& \textbf{0.434}
& 0.338
& 0.393
& 12.8\%
& \textbf{0.217}
& 0.121
& 0.163
\\
\phakg{}
& \textbf{0.587}
& 0.396
& 0.567
& 14.8\%
& \textbf{0.176}
& 0.086
& 0.165
\\
\gokg{}
& 0.571
& 0.523
& \textbf{0.601}
& 18.5\%
& 0.372
& 0.347
& \textbf{0.439}
\\
\bottomrule
\end{tabular}%
}
}
\end{table}

\begin{figure}[t]
\centering
\includegraphics[width=\columnwidth]{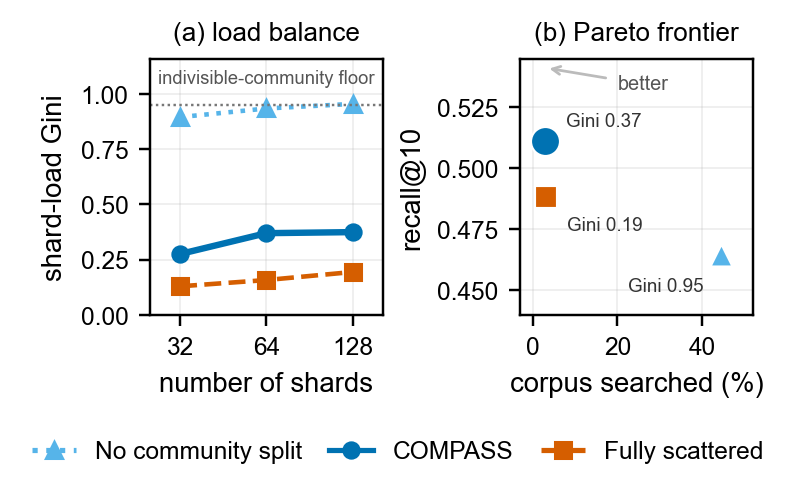}
\caption{Effect of community splitting on \prikg{}. Bounded splitting in \compass{} reduces imbalance caused by oversized communities without scattering all across many shards.}
\label{fig:frontier}
\end{figure}

\subsection{Bounded splitting balances load while retaining locality}
\label{sec:frontier}

\autoref{fig:frontier} isolates the effect of community splitting on \prikg{}. Keeping Louvain communities intact leaves the shard-load Gini near 0.95 because the largest community dominates one shard. Full scatter improves balance but fragments communities across shards, weakening selective retrieval. \compass{} instead splits only communities that exceed the shard-size bound.

At 128 shards (\autoref{tab:frontier}), no community split searches 44.6\% of the corpus with two shards. Full scatter reduces this to 3.1\% but lowers recall, whereas \compass{} searches 2.9\%, achieves the highest recall, and delivers the highest throughput. It lowers Gini to 0.37, while each original community spans only 1.02 shards on average. As the shard count increases from 32 to 128, the searched fraction falls from 7.4\% to 2.9\%, while related entities remain largely co-located.

\begin{table}[t]
\centering
\caption{Placement at 128 shards on \prikg{}. Community span is the average number of shards occupied by a community.}
\label{tab:frontier}
\small
\setlength{\tabcolsep}{4.5pt}
\resizebox{\columnwidth}{!}{%
\begin{tabular}{@{}lrrrrr@{}}
\toprule
Strategy
& Comm. span
& Gini
& Corpus searched
& Recall@10
& QPS \\
\midrule
No community split
& 1.00
& 0.95
& 44.6\%
& 0.465
& 15.0k \\
\textbf{\compass{}}
& 1.02
& 0.37
& \textbf{2.9\%}
& \textbf{0.511}
& \textbf{18.6k} \\
Fully scattered
& 1.78
& 0.19
& 3.1\%
& 0.488
& 16.2k \\
\bottomrule
\end{tabular}%
}
\end{table}

\subsection{Shard selection enables scale-out}
\label{sec:performance}

\begin{figure}[t]
\centering
\includegraphics[width=\columnwidth]{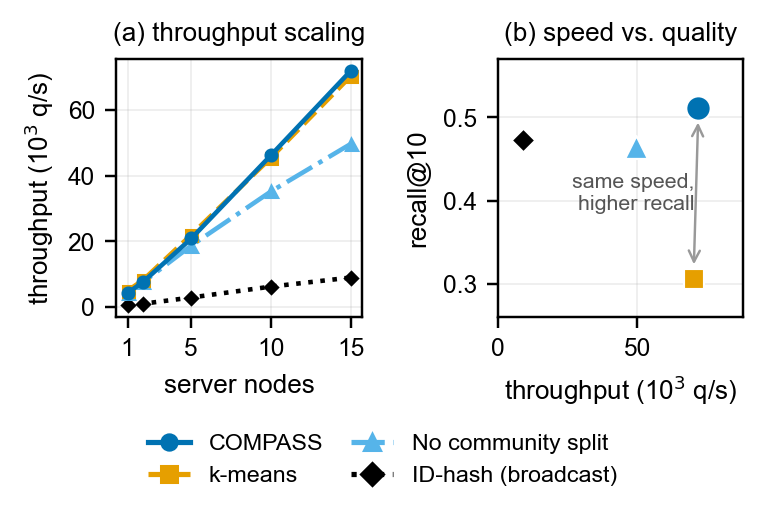}
\caption{Scaling on \prikg{}. (a) Selective methods benefit  more from additional nodes than broadcast. No community splitting scales less effectively because its selected shards remain large. (b) At 15 nodes, \compass{} and $k$-means provide similar throughput, but \compass{} achieves much higher recall.}
\label{fig:scaling}
\end{figure}

\autoref{fig:scaling} fixes the database at 60 shards and scales Qdrant from 1 to 15 nodes. Broadcast throughput increases with node count but remains $7$--$8\times$ below \compass{} and $k$-means, while its p95 latency stays near 8--10\,ms. At 15 nodes, \compass{} reaches 71.8k QPS at 1.03\,ms, compared with 9.0k QPS at 8.09\,ms for broadcast (\autoref{tab:sysres}). $k$-means achieves similar system performance (70.2k QPS at 0.99\,ms), confirming that selective search drives the speedup. However, placement determines retrieval quality: \compass{} achieves 0.511 recall, compared with 0.306 for $k$-means (\autoref{fig:scaling}(b)).

Resource measurements further explain these scaling results. Broadcast sends every query to all 60 shards and reaches 15.4\% peak CPU utilization, yet delivers only 9.0k QPS. By searching two small shards, \compass{} and $k$-means serve nearly $8\times$ more queries while keeping peak CPU utilization below 8\%. In contrast, no-community-split selects only two shards, but they contain about 45\% of the corpus. It therefore preserves 0.464 recall but reaches only 49.6k QPS with 1.73\,ms p95 latency. Its largest-shard nodes reach 17.7\% CPU utilization, exposing substantial load imbalance. Effective scale-out thus requires both selective routing and balanced shards. % sizes.

\begin{table}[t]
\centering
\caption{Query performance and peak resource use with 15 nodes and 60 shards. CPU is the busiest node’s utilization, sampled every 5\,s; memory is the largest per-node footprint, including Qdrant and in-memory shard data.}
\label{tab:sysres}
\small
\setlength{\tabcolsep}{4.5pt}
\resizebox{\columnwidth}{!}{%
\begin{tabular}{@{}lrrrrr@{}}
\toprule
Plan
& QPS (k)
& p95 (ms)
& Recall@10
& CPU (\%)
& Memory (GB) \\
\midrule
ID-hash (broadcast)
& 9.0
& 8.09
& 0.473
& 15.4
& 5.1 \\
$k$-means
& 70.2
& \textbf{0.99}
& 0.306
& 7.5
& 5.2 \\
No community split
& 49.6
& 1.73
& 0.464
& 17.7
& 6.1 \\
\textbf{\compass{}}
& \textbf{71.8}
& 1.03
& \textbf{0.511}
& 5.6
& 5.4 \\
\bottomrule
\end{tabular}%
}
\end{table}

\subsection{Overheads and configuration effects}
\label{sec:overhead}

\compass{} adds modest setup cost (\autoref{tab:buildcost}). Community detection, bounded splitting, and shard assignment take at most 1.21 s. Routing new data takes 1.4--20.2 s, and insertion plus HNSW indexing adds 4.3--8.2 s. Total construction time ranges from 6.0 to 29.6 s across the four KGs.

At run time, \compass{} places each new entity in $105~\mu\mathrm{s}$--$1.24$ ms and routes each query in $72~\mu\mathrm{s}$--$1.46$ ms. In terms of memory, the router holds only the entity-to-shard map, the truncated adjacency, and one centroid per shard: 4.6--13.1 MB in total, $29$--$79\times$ smaller than the whole fact embeddings it steers. The centroids come from a one-time scan of the vectors (9.5--38.5 s on \prikg{}, depending on cache state): a cost the $k$-means baseline pays for its core
routing rather than a fallback. Including this scan, the full \prikg{} build cost is amortized after about 0.75 million queries on 15 nodes.

\begin{table}[t]
\centering
\caption{\compass{} construction-time breakdown in seconds.}
\label{tab:buildcost}
\small
\setlength{\tabcolsep}{4.0pt}

\resizebox{\columnwidth}{!}{%
\begin{tabular}{@{}lrrrr@{}}
\toprule
Stage & \prikg{} & \hipkg{} & \phakg{} & \gokg{} \\
\midrule
Bulk placement
& 1.21 & 0.74 & 0.24 & 0.25 \\

\quad Community detection
& 0.15 & 0.10 & 0.06 & 0.07 \\

\quad Bounded splitting
& 0.80 & 0.61 & 0.05 & 0.04 \\

\quad Shard assignment
& 0.26 & 0.03 & 0.13 & 0.14 \\

New data routing
& 20.2 & 5.4 & 8.5 & 1.4 \\

Insertion (Qdrant)
& 1.1 & 0.8 & 0.5 & 0.3 \\

HNSW indexing
& 7.1 & 4.0 & 4.0 & 4.0 \\
\midrule
\textbf{End-to-end}
& \textbf{29.6}
& \textbf{10.9}
& \textbf{13.2}
& \textbf{6.0} \\
\bottomrule
\end{tabular}%
}
\end{table}

\begin{table}[t]
\centering
\caption{Effect of segments per shard (\prikg{}, 16 shards; \compass{} probes two shards). More segments leave recall flat but add a local search on every contacted shard; the effect compounds for broadcast, which contacts every shard.}
\label{tab:segments}
\small
\setlength{\tabcolsep}{5pt}
\resizebox{\columnwidth}{!}{%
\begin{tabular}{@{}rcccccc@{}}
\toprule
& \multicolumn{3}{c}{\compass{}}
& \multicolumn{3}{c}{Broadcast} \\
\cmidrule(lr){2-4}\cmidrule(lr){5-7}
Seg./shard & QPS (k) & p95 (ms) & R@10 & QPS (k) & p95 (ms) & R@10 \\
\midrule
2  & 16.0 & 1.4  & 0.491 & 8.5 & 2.3  & 0.471 \\
4  & 14.7 & 1.6  & 0.496 & 6.6 & 3.2  & 0.472 \\
8  & 12.1 & 1.9  & 0.498 & 3.3 & 8.3  & 0.477 \\
16 &  8.9 & 2.9  & 0.496 & 1.8 & 16.4 & 0.480 \\
\bottomrule
\end{tabular}%
}
\end{table}

Local index configuration has a larger effect on performance than graph routing (\autoref{tab:segments}). As the number of Qdrant segments per shard increases, recall remains flat, but each query must search more local indexes. \compass{} throughput falls by about half, while broadcast drops from 8.5k to 1.8k QPS and reaches 16.4 ms p95 latency. By contrast, increasing workers from 8 to 16 raises throughput from 28.9k to 54.9k QPS, with unchanged recall and peak node memory near 3\%. Balanced shards therefore benefit from added concurrency, whereas excessive segmentation adds unnecessary search work.

\subsection{When graph structure adds value}
\label{sec:value}

Graph-based placement is most useful when embedding similarity does not reflect KG links. To evaluate this, we remove graph edges and ask whether an entity embedding can retrieve the entity at the other end of each removed edge among its top-10 nearest neighbors. Hits@10 is only 1--2\% on \prikg{}, \hipkg{}, and \phakg{}, indicating that their embeddings rarely place linked entities close together. On these KGs, \compass{} improves recall over $k$-means by 0.161, 0.096, and 0.191, respectively. On \gokg{}, the embeddings recover about 50\% of the removed links, and the recall gain falls to 0.048.

\begin{figure}[t]
\centering
\includegraphics[width=0.92\columnwidth]{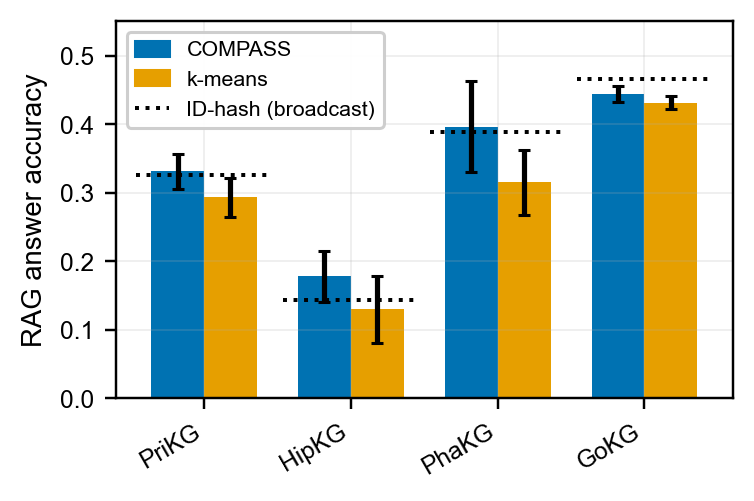}
\caption{Answer accuracy under the SciCUEval protocol, averaged across three language models. Improvements in retrieval quality carry through to downstream answers.}
\label{fig:rag}
\end{figure}

We next measure end-to-end RAG accuracy using Qwen2.5-7B, Llama-3.1-8B, and Gemma-2-9B (\autoref{fig:rag}). Averaged across the three readers, \compass{} outperforms $k$-means on every KG, improving accuracy from 0.293 to 0.331 on \prikg{}, 0.130 to 0.178 on \hipkg{}, and 0.315 to 0.396 on \phakg{}. On these KGs, where embeddings poorly preserve graph links, \compass{} also matches or exceeds broadcast; the largest gain is on \hipkg{} (0.178 versus 0.143). On \gokg{}, where embedding similarity already captures much of the graph structure, all methods perform similarly, with broadcast marginally ahead. These results indicate that improved placement translates into higher answer accuracy without changing the reader.

\section{Discussion and Limitations}
\label{sec:discussion}

\textbf{Data placement and routing are closely linked.}
Selective search reduces work by probing only a few shards, but its performance depends on both the number and size of the selected shards. The probe count affects total work, while the largest selected shard often shapes latency because the query must wait for all selected shards to finish. Simple layouts tend to favor one factor over the other. Keeping communities intact preserves locality and limits the probe set, but large communities can create oversized shards. Spreading communities more evenly improves balance   but may scatter related evidence across many shards. \compass{} seeks a middle ground by splitting oversized communities into bounded groups and routing queries over the resulting entity-to-shard map.

\textbf{The search schedule determines speed; placement determines recall.}
With the same two-shard budget, \compass{} and $k$-means have nearly identical throughput and tail latency because they search the same number of similarly sized shards. However, their recall differs significantly because $k$-means groups facts only by embedding similarity, while \compass{} keeps graph-related facts close when those relations are poorly captured by the embeddings. In other words, speed comes from balanced shards and a fixed probe count, while recall depends on which relationships the placement preserves. \compass{} changes data insertion and query routing rather than the ANN index itself, so it can be combined with better index structures or vector-compression methods.

\textbf{Graph structure is most useful when it complements embeddings.}
KG-guided insertion helps most when embedding similarity does not preserve the relations needed by the workload. Held-out edge recovery provides a simple way to estimate this complementarity before deployment: poor recovery suggests that the graph offers an independent signal for co-locating evidence, whereas strong recovery leaves less room for improvement. Our gold-edge ablation further suggests that the gain arises from persistent data organization rather than direct graph traversal at query time. The downstream RAG results also show that better placement improves LLM accuracy across the evaluated models.

\textbf{Limitations.}
Our evaluation covers modest-scale biomedical KGs, one embedding model, one vector database, and corpora of 44k--298k facts. Billion-scale, cross-domain,
and storage-backed deployments remain untested. \compass{} also benefits queries that can be reliably linked to the KG; unanchored queries fall back to centroid routing and receive no graph-based locality benefit. Finally, the bootstrap split and ground-truth edge ablation reduce leakage concerns but do not establish performance when relevant evidence is missing from the graph.

\section{Conclusion}
\compass{} reframes distributed vector search around how scientific data is organized before queries arrive. Rather than relying on embedding geometry alone, it uses the KG to group structurally connected facts, while subdividing only the largest groups to avoid uneven shard utilization. The result is a layout that supports selective search without dispersing the evidence needed for relational questions. Our evaluation also clarifies that efficiency and retrieval quality arise from different parts of the design. Searching fewer shards reduces system cost, but remains effective only when the underlying layout concentrates relevant data within those shards. \compass{} provides this concentration through KG-informed organization while preserving the existing ANN indexes within each shard. More broadly, our work shows that compact domain-specific structure can help distributed vector stores scale without making every query involve the entire database.

\section*{Acknowledgment}
This work was supported by the U.S. Department of Energy (DOE), Office of Science, Office of Advanced Scientific Computing Research, through the TIDES project, and by NSF grant 2411188. This research used resources of the Argonne Leadership Computing Facility at Argonne National Laboratory, under Contract No. DE-AC02-06CH11357.

\balance
\bibliographystyle{IEEEtran}
\bibliography{bibs/references}
\end{document}